\documentclass[sigconf]{acmart}
\usepackage{booktabs} 
\usepackage{kotex}
\usepackage{svg}
\usepackage{amsmath}
\usepackage{graphicx}
\usepackage{enumitem}
\usepackage{multirow}
\usepackage{threeparttable}
\usepackage{balance}
\usepackage{mathtools}
\usepackage{array}

\def\dsp{\def\baselinestretch{0.95507}\large\normalsize}
\dsp
\def\ffsp{\def\baselinestretch{0.95507}\large\normalsize}
\ffsp
\setcopyright{acmcopyright}

\acmDOI{10.1145/3605098.3636009}

\acmISBN{978-1-4503-8713-2/22/04}
\acmConference[SAC'24]{The 39th ACM/SIGAPP Symposium On Applied Computing}{April 8 –April 12, 2024}{Avila, Spain}
\acmYear{2024}
\copyrightyear{2024}

\acmArticle{4}
\acmPrice{15.00}

\begin{document}
\title{RecKG: Knowledge Graph for Recommender Systems}
  
\renewcommand{\shorttitle}{RecKG: Knowledge Graph for Recommender Systems}

\author{Junhyuk Kwon}
\affiliation{%
  \institution{Inha University}
  \streetaddress{P.O. Box 1212}
  \city{Incheon} 
  \state{South Korea} \\
  \postcode{43017-6221}
}
\email{tree.jhk@gmail.com}

\author{Seokho Ahn}
\affiliation{%
  \institution{Inha University}
  \streetaddress{P.O. Box 1212}
  \city{Incheon} 
  \state{South Korea} \\
  \postcode{43017-6221}
}
\email{sokho0514@gmail.com}

\author{Young-Duk Seo}
\authornote{Corresponding author}
\affiliation{%
  \institution{Inha University}
  \streetaddress{P.O. Box 1212}
  \city{Incheon} 
  \state{South Korea} \\
  \postcode{43017-6221}
}
\email{mysid88@inha.ac.kr}

\renewcommand{\shortauthors}{Kwon et al.}

\begin{abstract}

Knowledge graphs have proven successful in integrating heterogeneous data across various domains. 
However, there remains a noticeable dearth of research on their seamless integration among heterogeneous recommender systems, despite knowledge graph-based recommender systems garnering extensive research attention. 
This study aims to fill this gap by proposing RecKG, a standardized knowledge graph for recommender systems. 
RecKG ensures the consistent representation of entities across different datasets, accommodating diverse attribute types for effective data integration. 
Through a meticulous examination of various recommender system datasets, we select attributes for RecKG, ensuring standardized formatting through consistent naming conventions. 
By these characteristics, RecKG can seamlessly integrate heterogeneous data sources, enabling the discovery of additional semantic information within the integrated knowledge graph. 
We apply RecKG to standardize real-world datasets, subsequently developing an application for RecKG using a graph database. 
Finally, we validate RecKG's achievement in interoperability through a qualitative evaluation between RecKG and other studies. 
Implementation of RecKG can be found at \href{https://github.com/tree-jhk/RecKG}{\textcolor{blue}{https://github.com/tree-jhk/RecKG}}.

\end{abstract}

%
%



\maketitle

\section{Introduction}


Recommender systems generally encounter the data sparsity issue due to insufficient interactions between users and items \cite{yang2023knowledge, kgcl, mcclk}. To address this concern, studies in recommender systems have utilized knowledge graphs (KG), which contain valuable auxiliary information about users and items through relational connections \cite{kgcn, kgin, krdn}. KG-based recommender system enhance the quality of user and item representations by leveraging diverse entities and relations extracted from the KG. Furthermore, they enable the provision of explanations for the recommendation outcomes by analyzing the connections among users, items, and supplementary information. Given these characteristics and merits, numerous KG-based recommender systems have emerged. Most of these studies aim to enhance the quality of embedding representations \cite{kprn, kgat, kgcn} or focus on addressing data quality concerns \cite{kgcl, krdn}. Their goals is to leverage KG to improve recommender system performance and navigate data efficiently.

Although various studies on KG-based recommender systems are ongoing, there has been limited study on the integration of KG between recommender systems. Integrating KG results in the broadening of the existing knowledge graph, incorporating attributes from disparate recommender systems \cite{kddzzang}. Furthermore, achieving interoperability for KG facilitates seamless integration, commencing knowledge enrichment by drawing information from other KG and broadening the scope of entities and relations. This approach maximizes the two core benefits of KG-based recommender systems: alleviating data sparsity and enhancing explainability \cite{kddzzang, kprn}. 

However, integrating KGs among recommender systems to achieve interoperability can be challenging due to diverse semantic representations across different systems \cite{ jirkovsky2016understanding, evkg, linkclimate, urbankg}. Addressing two key aspects, namely, \textit{consistency} and \textit{diversity}, is crucial to overcoming these challenges. To begin with, the failure to maintain consistency occurs when different representations are employed to model the same concept \cite{evkg, grangel2021analyzing, jirkovsky2016understanding}. In such cases, confusion and ambiguity may arise when attempting to identify information. Additionally, the inability to ensure diversity arises when different items are missing from distinct data sources \cite{grangel2021analyzing}. In such scenarios, there is a risk of failure to capture crucial information.

In this study, we propose RecKG, a standardized KG for recommender systems, to address hindrances in KG interoperability, particularly concerning \textit{consistency} and \textit{diversity}. RecKG is composed of entities essential for recommender systems, with a primary focus on users and items. Through RecKG, we aim to resolve the following issues. First, achieving consistency in data integration is essential to ensure that data from different systems uniformly represent identical concepts and attributes. RecKG tackles this challenge by adopting comprehensive naming conventions for attributes associated with users and items across various recommender systems. Second, during the process of data integration, RecKG must cover a wide range of attributes to secure diversity within the recommender system research area. To accomplish this, RecKG is designed to minimize the occurrence of missing attributes within the KG tailored for recommender systems and enable broad coverage of attributes across various recommender systems.

In summary, RecKG enhances consistency and diversity within the knowledge graph (KG) for recommender systems, resulting in seamless integration and improved recommendation outcomes. We outline our key contributions as follows:
\begin{itemize}[noitemsep, leftmargin=1.2em]
    \item
        We propose RecKG, a standardized KG for recommender systems, to achieve interoperability.
    
    \item
        We provide detailed explanations for the attributes that constitute RecKG and their relevance in modeling user-item interactions based on the diagram of RecKG.
        
    \item
        Standardizing two real-world datasets based on RecKG, we apply it to a graph database management system. In doing so, we verify the interoperability of RecKG and conduct a qualitative evaluation.
\end{itemize}
\begin{figure*}
  \centering
  \includegraphics[width=0.9\linewidth]{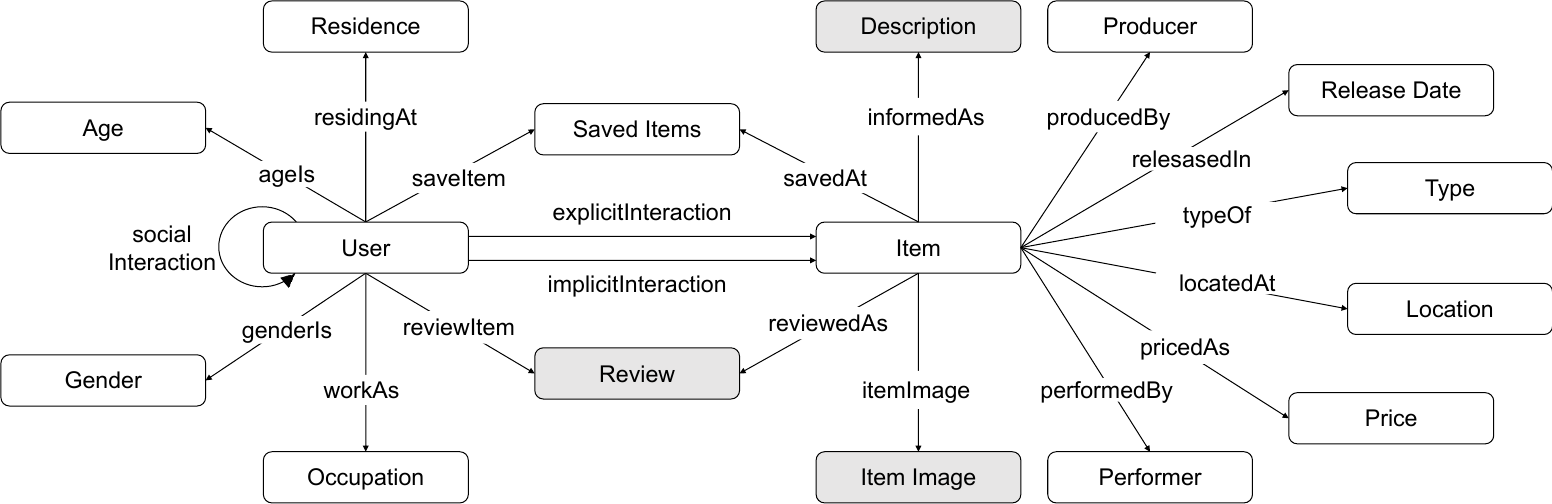}
  \caption{The diagram of RecKG. Nodes represent user and item attributes, with gray nodes indicating text or image attributes. Edges denote relations between nodes.}
  \label{fig:RecKG}
\end{figure*}

\section{Related Works}
In this section, we first elucidate the benefits of integrating KG across various domains. Subsequently, we introduce existing studies on KG-based recommender systems.

\subsection{Knowledge Graph Integration Across Diverse Fields}
In various research fields, adopting KG has proven effective for the seamless integration of relevant data. Consolidating datasets through KG, we can facilitate enhanced data-driven decision-making and data analysis. Below, we present three distinct studies of KG-based integration in other research fields, not related to recommender systems.


In the field of climate research, LinkClimate \cite{linkclimate}, an interoperable KG platform for climate data, is proposed. Through LinkClimate, the integration of diverse climate data using KG has been explored, enhancing the understanding of climate interactions. This research enables cross-domain analysis among various climate data. 

In the field of Electric Vehicles (EV), EVKG \cite{evkg}, comprising EV-related ontology modules from various sources, is proposed. This integration enhances interoperability with other KG, simplifying the complex EV ecosystem and supporting decision-making related to technology and infrastructure. 

Finally, UrbanKG \cite{urbankg}, an urban KG system that leverages KG for the integration of diverse urban datasets, enhancing data analysis efficiency in the field of urban computing, is introduced. A structured schema for UrbanKG is presented, offering a systematic representation of fundamental urban entities and their interconnections. This schema encompasses a wide range of relations for defining connections between entities, handles spatial-temporal urban data, and seamlessly integrates external knowledge bases.


The data integration serves to diversify data and acquire relationship information between different systems. However, research in the three fields mentioned earlier all faces a common challenge, which is the integration of non-standardized heterogeneous data \cite{linkclimate, evkg, urbankg}. This challenge involves managing the diversity of data sources and structures from different systems. The ultimate goal of our study is to standardize a KG composed of attributes and relations suitable for recommender systems. Through RecKG, we represent various recommender system datasets in a standardized format and contribute to their seamless integration.

\subsection{Knowledge Graph-based Recommender System}
In the KG-based recommender systems, many studies aim to utilize KG containing auxiliary information for items and users, addressing the data sparsity problem. To effectively model this auxiliary information from the KG, KPRN \cite{kprn} generates path representations for both entities and relations within a KG, utilizing them to provide recommendations for users. RuleRec \cite{rulerec} is a rule-based recommender system that extracts understandable rules from a KG, consolidating the derivation of explainable rules from the KG.

Furthermore, there are ongoing studies focused on efficiently modeling KG with rich information within neural network structures.
KGCN \cite{kgcn} is a framework using graph convolutional networks (GCN) to create embeddings for items in a KG and is renowned for its effectiveness in mining important attributes and details within a KG.
KGAT \cite{kgcn} is an attention-based recommender system that captures high-order structure and semantic information from a KG.
MCCLK \cite{mcclk} introduces a multi-level cross-view contrastive learning mechanism for KG-based recommender systems, incorporating three distinct graph views: the user-item graph as a collaborative view, the item-entity graph as a semantic view, and the user-item-entity graph as a structural view. 

Several studies are also focused on mitigating the impact of noise in KG-based recommender systems, with the goal of improving their overall performance.
A KG augmentation scheme is introduced \cite{kgcl}, aimed at suppressing KG noise during information aggregation to address the challenge of the long-tail distribution of entities in KG and mitigate information noise between items and entities in KG-enhanced recommender systems. KRDN \cite{krdn} is proposed to address challenges stemming from the unnecessary spread of knowledge and noise in user-item interactions. KRDN efficiently trims irrelevant knowledge connections, eliminates disruptive implicit feedback, and utilizes an adaptive knowledge refinement approach to preserve high-quality KG triplets.

Most of existing studies primarily focus on constructing a single KG for the recommender system. On the other hand, research on multiple KG-based recommender systems is still relatively inadequate. Integrating KG from different recommender system datasets is feasible due to the presence of shared common items within the same domain (e.g., the movie domain). In some studies, researchers augment a single recommendation system dataset with an external knowledge base to compensate for the lack of information in the dataset. For instance, KPRN \cite{kprn} combines the ML dataset with IMDb\footnote{https://www.imdb.com/} dataset using movie titles and release dates. In KGCN \cite{kgcn}, the ML dataset is linked to Microsoft Satori\footnote{https://searchengineland.com/library/bing/bing-satori} using matching Satori IDs to construct a KG. The addition of attributes from different systems expands the existing KG \cite{kddzzang}. However, KG integration is a complex task because, even within the same domain, KG developed in different systems typically has structural disparities of data representation. Existing studies use predesigned KGs for integration between specific datasets, with limited attention to the generalization and standardization for KG integration \cite{remr, kddzzang}.
\section{Knowledge Graph for Recommender Systems}
In this section, we introduce RecKG, a standardized KG designed for recommender systems. We first articulate the goals of RecKG, focusing on achieving interoperability. Subsequently, we present the components of RecKG, including attributes. At the end of this section, we provide a practical demonstration of RecKG's application on real-world datasets, emphasizing the modeling of the integrated KG using RecKG and showcasing its applicability.

\subsection{Goal of RecKG}\label{goal of reckg}
While the majority of substantial studies concentrate on enhancing the performance of KG-based recommender systems, there is a relative scarcity of research on KG integration. Even existing studies on integrated KG-based recommender systems often assume that KG integration is preconfigured, neglecting the consideration of how to integrate different KG \cite{remr, kddzzang}. 

Integration of KG from different systems becomes feasible owing to the presence of shared common items within the same domain, such as the movie domain. When these disparate systems seamlessly integrate, \textit{interoperability} is attained. According to IEEE, interoperability is defined as ``The ability of two or more systems or components to exchange information and and to use the information that has been exchanged \cite{institute1990ieee}.'' The challenge in achieving interoperability arises from varying representations of same entities across diverse systems. To address this challenge, we introduce RecKG, a standardized KG for recommender systems, composed of entities essential for recommender systems, with a primary focus on users and items. RecKG tackles this challenge by emphasizing two key aspects: \textit{consistency} and \textit{diversity}, with the overarching goal of achieving interoperability.


\begin{table*}[]
\caption{User-item interactions table}
\label{tb:user-item interaction table}
\resizebox{0.95\textwidth}{!}{  
\begin{threeparttable}
\setlength\tabcolsep{10pt}
\begin{tabular}{ccccc}
\toprule
\textbf{Interaction} & \textbf{Interactive Behavior Type} & \textbf{Rating Type}   & \textbf{Datasets}   \\
\midrule
\multirow{2}{*}{\textbf{Explicit}}  & Dislike/Neutral/Like & Ordinal & Yelp\tnote{1}\\  
& Rating & Interval & Yelp , Gowalla\tnote{1} , Amazon\tnote{2} , MovieLens\tnote{3}\\ 
\midrule
\multirow{3}{*}{\textbf{Implicit}} & Click, Purchase                    & Unary          & Taoba\tnote{2} , Tmall\tnote{2} , BeiBei\tnote{2} , IJCAI\tnote{2} \\  
                                       & Bookmark, Cart, Favorite, Playlist & Unary           & Yelp , Taobao , Tmall , BeiBei , IJCAI , Last.FM\tnote{4} \\ 
 & Comment, Review, Tip               & Text       & Yelp , Gowalla , Amazon , E-commerce\tnote{2} , MovieLens\\ 
\bottomrule
\end{tabular}
\begin{tablenotes}
\item[1] POI $^2$ E-commerce $^3$ Movie $^4$ Music
\end{tablenotes}
\end{threeparttable}
}
\end{table*}

\begin{itemize}[noitemsep, leftmargin=1.2em]
    \item \textbf{\textit{Consistency}}\textbf{:}
Conflicts in interoperability emerge when different representations are employed to model identical entities \cite{grangel2021analyzing}. In such cases, ambiguity may arise when attempting to identify information. Therefore, for effective data integration, it becomes imperative to ensure that data originating from other systems uniformly represent same entities. RecKG addresses this challenge by adopting a comprehensive naming scheme for attributes associated with users and items across diverse recommender systems. This approach mitigates discrepancies in entity names when referring to the same entity \cite{jirkovsky2016understanding}. This standardization ensures \textit{consistency} among heterogeneous datasets.

\item \textbf{\textit{Diversity}}\textbf{:}
Interoperability conflicts can also arise when different items are absent in distinct data sources \cite{grangel2021analyzing}. This conflict is a crucial consideration when determining the attributes to include in RecKG. This is because, when standardizing a recommender system dataset using RecKG, it should encompass all the vital attributes for recommender systems. Thus, RecKG must be designed to incorporate diverse attributes that can influence the modeling of potential interactions between users and items. To achieve this goal, we meticulously examined attributes from various recommender system datasets \cite{musicrec, poirec, kgrec, foursquare, gowalla, residence, multimodalkg, user_behavior, xia2020multiplex} to include all those that are instrumental in understanding user-item preferences.
\end{itemize}

\subsection{Attributes of RecKG}\label{sec:Attributes in the RecKG}

RecKG is comprised of fundamental attributes of users and items, drawn from various datasets of recommender systems \cite{musicrec, poirec, kgrec}. In KG-based recommender systems, the attributes of users and items are considered as entities alongside users and items themselves. RecKG includes structured and named entities to encompass a wide range of non-standardized attributes. The standardized entities within RecKG are shown in Figure \ref{fig:RecKG}, represented as nodes.

\subsubsection{Attributes of Users}
Users represent individuals utilizing the recommender system. The attributes of users encompass demographic information, which serves as valuable data to distinguish users effectively within recommender systems and mitigate cold-start issues \cite{demographic_rec2, demographic_coldstart, demographic_rec1}. Most demographic user profiles typically include age, gender, and occupation. Additionally, we have integrated user residence information into RecKG attributes, considering the Point of Interest (POI) domain \cite{foursquare, gowalla}.

In RecKG, entities corresponding to attributes of \textit{user} include the following:

\begin{itemize}[noitemsep, leftmargin=1.2em]
    \item \textbf{\textit{Age}}\textbf{:}
        Since \textit{age} is a numerical data, we categorize it into age groups to treat it as categorical information. 
        This simplification offers advantages because similar age groups often share common interests. In recommender system research that utilizes demographic information \cite{demographic_rec1, demographic_rec2}, users are appropriately divided into age groups through binning techniques to align with the specific domain.
    
    \item \textbf{\textit{Gender}}\textbf{:}
        \textit{Gender} is a binary attribute, typically categorized as male or female \cite{demographic_rec1}. However, when combined with other demographic information, it enhances user differentiation, resulting in more distinct user profiles, such as `A Spanish woman in her 50s' or `A Korean student in his 20s.'
    
    \item \textbf{\textit{Residence}}\textbf{:}
        In the field of POI recommendation, there are cases where user residence information is provided. Therefore, some studies leverage user residence information \cite{residence, he2018personalized}. Depending on the dataset, this information may be provided in a precise manner, such as latitude and longitude, or indirectly, by country or state. Hence, \textit{residence} is grouped according to the given data format and utilized as an attribute in RecKG.
    
    \item \textbf{\textit{Occupation}}\textbf{:}
        Occupational information for users is provided in broad categories, such as artist and engineer. A study revealed that users' preferences vary significantly based on occupation \cite{occupation}. Therefore, to group users accordingly, we adopted \textit{occupation} as an attribute in RecKG.
\end{itemize}

\subsubsection{Attributes of Items}
Items in recommender systems can vary depending on the domain, such as movies, music, E-commerce, POI, and more. While attributes of the \textit{user} entity do not vary across the domains, the \textit{item} entity can have significantly different attributes. Consequently, it is essential to carefully choose attributes of the item entity that are adaptable and comprehensive to diverse datasets while ensuring naming consistency in RecKG.

\begin{itemize}[noitemsep, leftmargin=1.2em]
    \item \textbf{\textit{Performer}}\textbf{:}
        A \textit{performer} refers to an individual or a group that plays a specific role or performs. Specific examples of movie and music recommender systems are as follows: In the movie domain, a performer entity refers to the actor or the actress in a movie. These individuals typically play the main characters, and their acting skills and popularity significantly influence the quality of the movie. In the music domain, the performer entity refers to singer or musician. Their style of song or music greatly influences a user's preferences. In essence, the performer entity embodies the central figure or participant of the work of art within each domain, and their role and technical/artistic contributions hold a significant position in recommendations.
        
    \item \textbf{\textit{Producer}}\textbf{:}
        The attribute \textit{producer} refers to the entity responsible for planning, creating, distributing, and managing items across various domains. Depending on the domain, the specific definitions and examples of a producer entity are as follows: In the movie domain, a producer plays a pivotal role in the film production process, which encompasses budgeting, planning, and overseeing production. This role can be further specified, such as a director or writer. In the music domain, a producer is actively engaged in recording and producing music, collaborating with music artists to plan and create music. This role can also be specified, such as music producer or entertainment. It is worth noting that the specific entities or organizations corresponding to the producer attribute may vary depending on the domain or dataset. For that reason, the producer could represent many attributes in a single dataset, they can be denoted as \textit{Producer}\textsubscript{i} (\(1 \leq i \leq n\)), where $n$ corresponds to the number of attributes associated with producer. For instance, when converting a specific dataset into RecKG and there are multiple attributes corresponding to producer entity, such as art directors, directors, and writers, each of them can be represented as \textit{Producer}\textsubscript{1}, \textit{Producer}\textsubscript{2}, \textit{Producer}\textsubscript{3}.
    
    \item \textbf{\textit{Type}}\textbf{:}
        The attribute \textit{type} represents a specific category, genre, or classification of an item, used to identify the type to which the item belongs. Depending on the domain, the definition of type can be further specified as follows: In the movie domain, the type entity indicates the genre of the film, encompassing categories that categorize movies. In the music domain, the type entity signifies the genre of the music, including styles like rock, pop and jazz. In the E-commerce domain, the type entity represents the category or sub-category of products like appliances, clothing and electronics. In the POI domain, the type entity denotes the category of a place, including venue categories like hotels, restaurants and tourist attractions. In essence, the type entity  categorize items based on similar characteristics in recommender system.
    
    \item \textbf{\textit{Description}}\textbf{:}
        \textit{Description} refers to textual information that provides detailed information about an item. The definition of description can be elaborated as follows: In the movie domain, it represents the plot or storyline of a film. In the E-commerce domain, it provides detailed product explanations. Within KG-based recommender system modeling, incorporating text data into the KG can improve the quality of recommendation \cite{multimodalkg}.
    
    \item \textbf{\textit{Item Image}}\textbf{:}
        \textit{Item Image} is an image attribute that represents the visual depiction of an item, used in recommender systems. In the movie domain, it corresponds to the movie poster. In the music domain, it signifies the album cover image, while in the E-commerce domain, it represents the product image. In the POI domain, it denotes pictures of places or tourist attractions. Research on multi-modal KG-based recommender systems has demonstrated that leveraging image data with KG supplements the quality of recommendation \cite{multimodalkg}.
    
    \item \textbf{\textit{Location}}\textbf{:}
        Most POI recommendation studies leverage location-nased social network (LBSN) \cite{seo2021point, long2023decentralized}. Diverse LBSN datasets provide location information, either explicitly with longitude and latitude, or indirectly, such as through the country or state. The \textit{location} entity aligns with such information from LBSN.
    
    \item \textbf{\textit{Price}}\textbf{:}
        \textit{Price} is a critical factor that significantly impacts a user's decision when making product purchases in the e-commerce domain \cite{price}. This quantitative attribute for products is sometimes discretized into a number of bins, with each product encoded to represent the respective interval to which it belongs \cite{chen2014does}.
    \item \textbf{\textit{Release Date}}
        \textit{Release date} represents the data an item was released. This attribute is crucial for profiling user preferences, especially identifying users who tend to favor the latest items.
\end{itemize}


\subsubsection{User-Item Interactions}
In KG-based recommender systems, a collaborative KG (CKG) is utilized as a unified relational graph representing the relations between items and attributes, along with user-item interactions \cite{kgat}. This approach combines knowledge-based information with collaborative filtering effects into the KG through the utilization of the item-attribute and the user-item interaction. Table \ref{tb:user-item interaction table} summarizes the user-item interactions in representative recommender system datasets.

\begin{figure}
  \centering
  \includegraphics[width=1.0\linewidth]{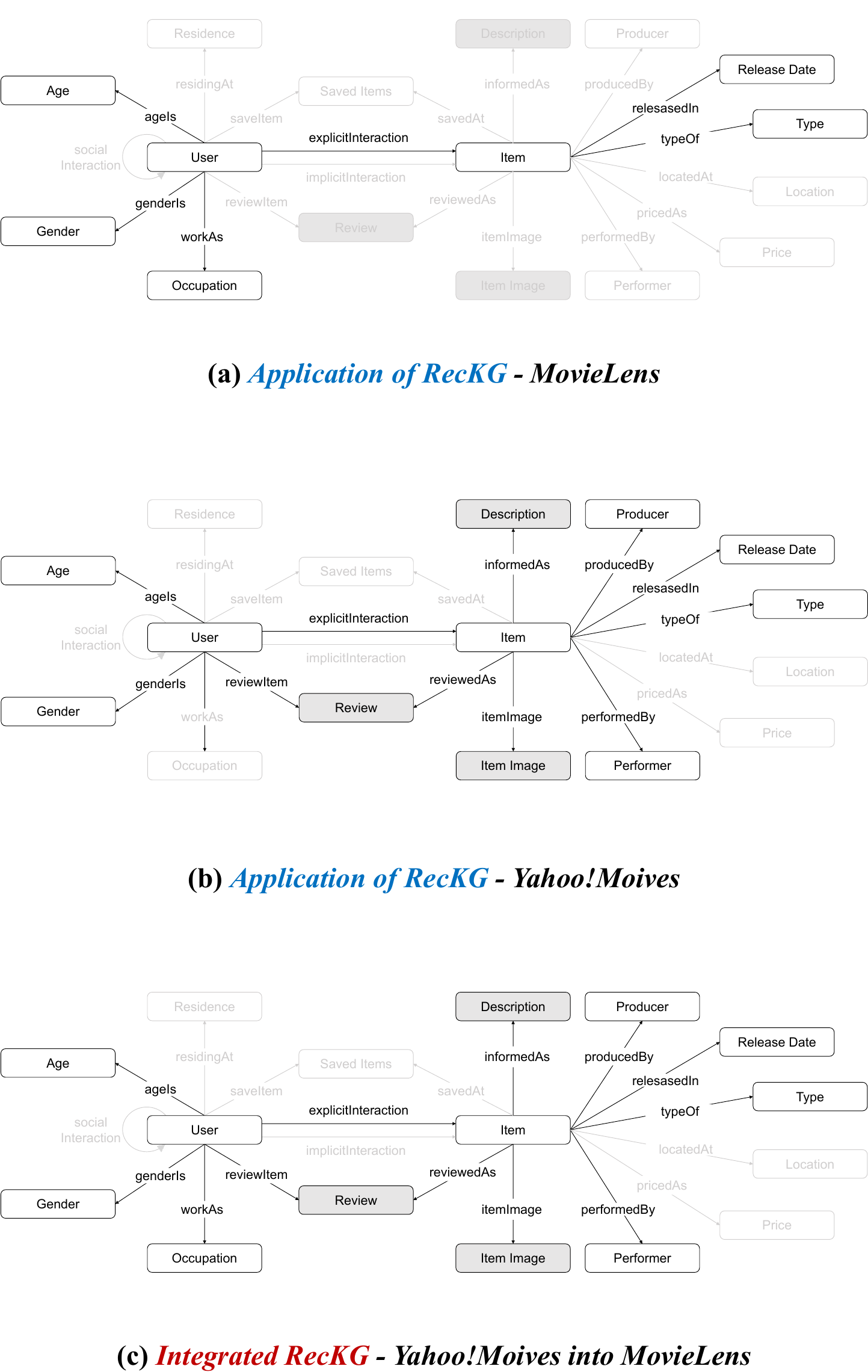}
  \caption{Design based on RecKG. (a) and (b) show KG designs for the MovieLens and Yahoo!Movies datasets using RecKG, respectively. (c) illustrates the integrated KG design, obtained by combining the KG designs presented in (b) into (a).}
  \label{fig:figure2}
\end{figure}

\begin{itemize}[noitemsep, leftmargin=1.2em]
    \item \textbf{\textit{explicitInteraction}}\textbf{:}
        The \textit{explicitInteraction} relation refers to direct and explicit interactions in which users provide feedback or preference information on items. Examples of such interactions include interval-based ratings (e.g., numerical scores between 1 and 5), binary ratings (e.g., Dislike or Like), or ordinal ratings (e.g., Dislike, Neutral, or Like) \cite{recsys_textbook}. In several KG-based recommender systems, interval-based ratings are often binarized by considering ratings above a certain threshold as positive feedback \cite{multi_vae}. In the case of binary ratings, items marked as `like' or `good' are commonly interpreted as positive feedback, and only positive feedback is utilized as relations between users and items in the CKG, while negative feedback is not \cite{wang2018ripplenet, kgat}.
    
    \item \textbf{\textit{implicitInteraction}}\textbf{:}
        The \textit{implicitInteraction} relation refers to indirect and implicit feedback provided by users to items. Examples of implicit interactions are clicks and purchases, which signify a user's preference for an item, but do not provide information on dislikes \cite{recsys_textbook}. For a specific user, items that have not been interacted with are considered as negative interactions, though in practice, users interact with only a small fraction of items within a system. Thus, a substantial number of items are treated as negative interactions. Due to this, many recommender systems based on implicit feedback, including KG-based recommender systems \cite{kgat, kprn}, employ the negative sampling technique \cite{ncf, youtube}. This technique efficiently optimizes the loss function of models that predict interactions by sampling only a small fraction of items with which the user has not interacted and incorporating them into the loss function.
        
    \item \textbf{\textit{Saved Items}}\textbf{:}
        \textit{Saved Items} are a part of implicit interactions, where users store items they personally prefer in dedicated storage. Depending on the system, these items can be stored separately in locations like playlists or in common storage options such as add-to-carts or add-to-favorites for archival purposes. In addition, the saved items entity can offer insights for profiling users' personalized interests in items, in contrast to purchases, which might be delayed and do not occur immediately \cite{user_behavior}.    
    
    \item \textbf{\textit{Review}}\textbf{:}
        \textit{Review} is a textual expression of user preferences for items. In some studies on recommender systems \cite{textreview_kg, textreview, textreview_survey, qiu2021u}, textual data is transformed into embedding vector, which are subsequently integrated into recommender systems as supplementary features.
        
\end{itemize}

\begin{figure}
  \centering
  \includegraphics[width=0.848\linewidth]{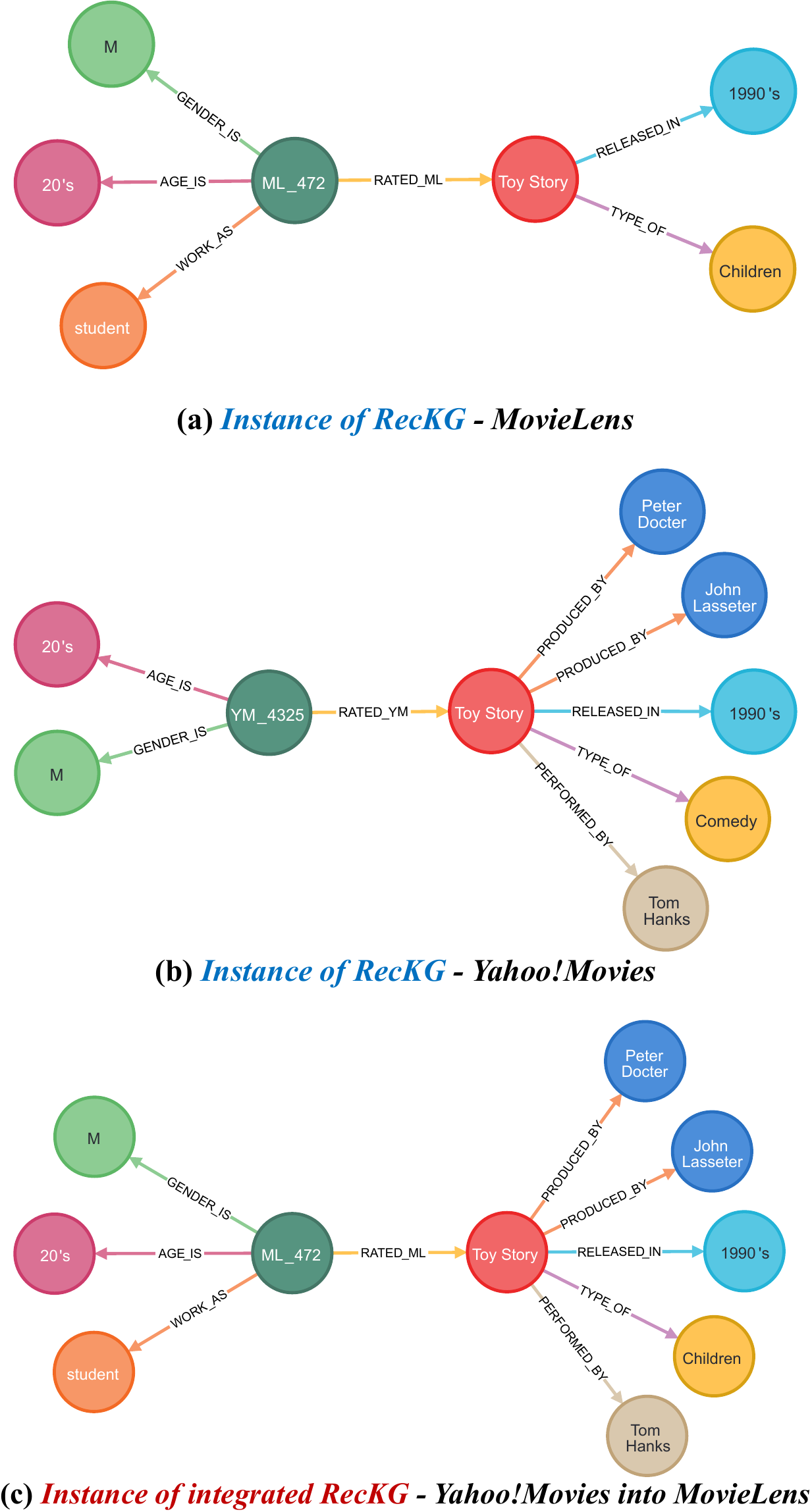}
  \caption{Instances of RecKG. (a) and (b) show sample subgraphs from the MovieLens and Yahoo!Movies datasets in RecKG. (c) illustrates a sample subgraph when integrating Yahoo!Movies into MovieLens in RecKG.}
  \label{fig:figure3}
\end{figure}

\subsection{Design based on RecKG}
The KG for recommender systems, represented as Figure \ref{fig:RecKG}, has been designed to encompass various domains and attributes, as explained in Section \ref{sec:Attributes in the RecKG}. Accordingly, RecKG is tailored to standardize a wide range of real-world datasets, covering a broad scope. Due to this comprehensive nature, most attributes in real-world recommender system datasets can be represented through the entities in RecKG. Figure Figure \ref{fig:figure2} (a) and (b) illustrate the KG designs for real-world datasets, specifically the ML 100k\footnote{https://grouplens.org/datasets/movielens/100k/} and the Yahoo!Movies\footnote{https://webscope.sandbox.yahoo.com/} (YM) datasets, standardized with RecKG. The blurred portions denote attributes that do not exist in the particular dataset. Further details about the datasets are provided in Section \ref{sec:Instance of RecKG}. 

In comparison to the YM, the ML contains fewer attributes. The limited number of attributes in the ML can only capture a small subset of the various reasons for user preferences regarding items. To maximize the advantages of KG-based recommender systems, such as explainability, it is essential to incorporate a more diverse set of attributes for the item entity. To complement the limited data in the ML, we integrated it with the YM, which contains fruitful information, as shown in Figure \ref{fig:figure2} (c).

In general, enhancing a KG that initially contains limited information with external knowledge bases can mitigate data sparsity. Figure \ref{fig:figure2} (c) illustrates the integrated RecKG of the YM into the ML. As in prior studies \cite{kprn, kgcn}, we have incorporated information from the YM into the ML, presenting the design of an augmented ML. The YM contains various attributes, alongside image (i.e., the item image attribute) and text (i.e., the review and the description attribute) information, enabling the potential for cross-modal modeling, as investigated in research on multi-modal KG-based recommender systems \cite{multimodalkg}. Through this integration, a recommender system using an augmented KG gains the capability to alleviate data sparsity and conduct a broader spectrum of reasoning.

\section{Case Study and Evaluation}
In this section, we demonstrate the application of RecKG to various real-world recommender system datasets using the graph database management system Neo4j\footnote{https://neo4j.com/}. This not only showcases the necessity of RecKG for KG-based recommender systems but also illustrates its implementability using a graph database. Finally, we delve into the exploration of additional semantics acquired through the seamless integration of KG facilitated by RecKG.

\subsection{Integration of Real-World Dataset}\label{sec:Instance of RecKG}
\subsubsection{Instance of RecKG}
Real-world recommender system datasets can be standardized using RecKG. In Figure \ref{fig:figure3} (a) and (b), we have concretely instantiated the ML-100k dataset and the YM dataset into actual instances in a graph database according to the schema of RecKG. Figure \ref{fig:figure3} (a) represents a sample subgraph extracted from the ML dataset, which includes user, movie, gender, age, occupation, release date, and genre. When applying the item's side information to Figure \ref{fig:RecKG}, these attributes are mapped as follows: The release date corresponds to the released date attribute, while genre corresponds to the type attribute. Figure \ref{fig:figure3} (b) illustrates a sample subgraph extracted from the YM dataset, which includes user, movie, gender, age, release year of item, director, cast, genre, and more. When integrating the item's side information into Figure \ref{fig:RecKG}, the attributes are mapped as follows: The release year maps to the released date attribute, genre corresponds to the type attribute, cast becomes the performer attribute, director map to the producer attribute. 

Instances applied in RecKG's design for the ML dataset and the YM dataset demonstrate that the adopted attribute names of RecKG effectively and consistently represent attributes in real-world datasets. This highlights the achievement of one of RecKG's goals, \textit{consistency}. Additionally, all attributes from both datasets are comprehensively represented using RecKG's adopted attribute names, successfully attaining \textit{diversity}.

\subsubsection{Integration of RecKG}
To validate RecKG's impact on achieving the \textit{interoperability} between two KG for recommender systems, we confirm the integrated scenario involving the ML dataset and the YM dataset. To integrate the YM dataset into the ML dataset, we embedded the attributes of YM dataset's items, which are movies, into ML dataset's items where both the title and release year are identical. It is worth noting that since both datasets are structured based on RecKG, the attribute names used are consistently identical. Therefore, thanks to this \textit{consistency}, the attribute names remain unchanged in the integrated RecKG without any modifications, as shown in Figure \ref{fig:figure3} (c).

As a result of this integration, the KG expands with the addition of more entities compared to its state before expansion. This extended KG contains a wealth of semantic information compared to the original KG, enabling more diverse recommendations along with more detailed explanations. As shown in Figure \ref{fig:figure3}, several nodes have been added from the YM dataset, specifically the performer and producer nodes. This expanded RecKG enhances the depth of semantic understanding through the inclusion of these additional nodes, enabling the delivery of more nuanced recommendation results to users.

\begin{figure}
  \centering
  \includegraphics[width=0.9\linewidth]{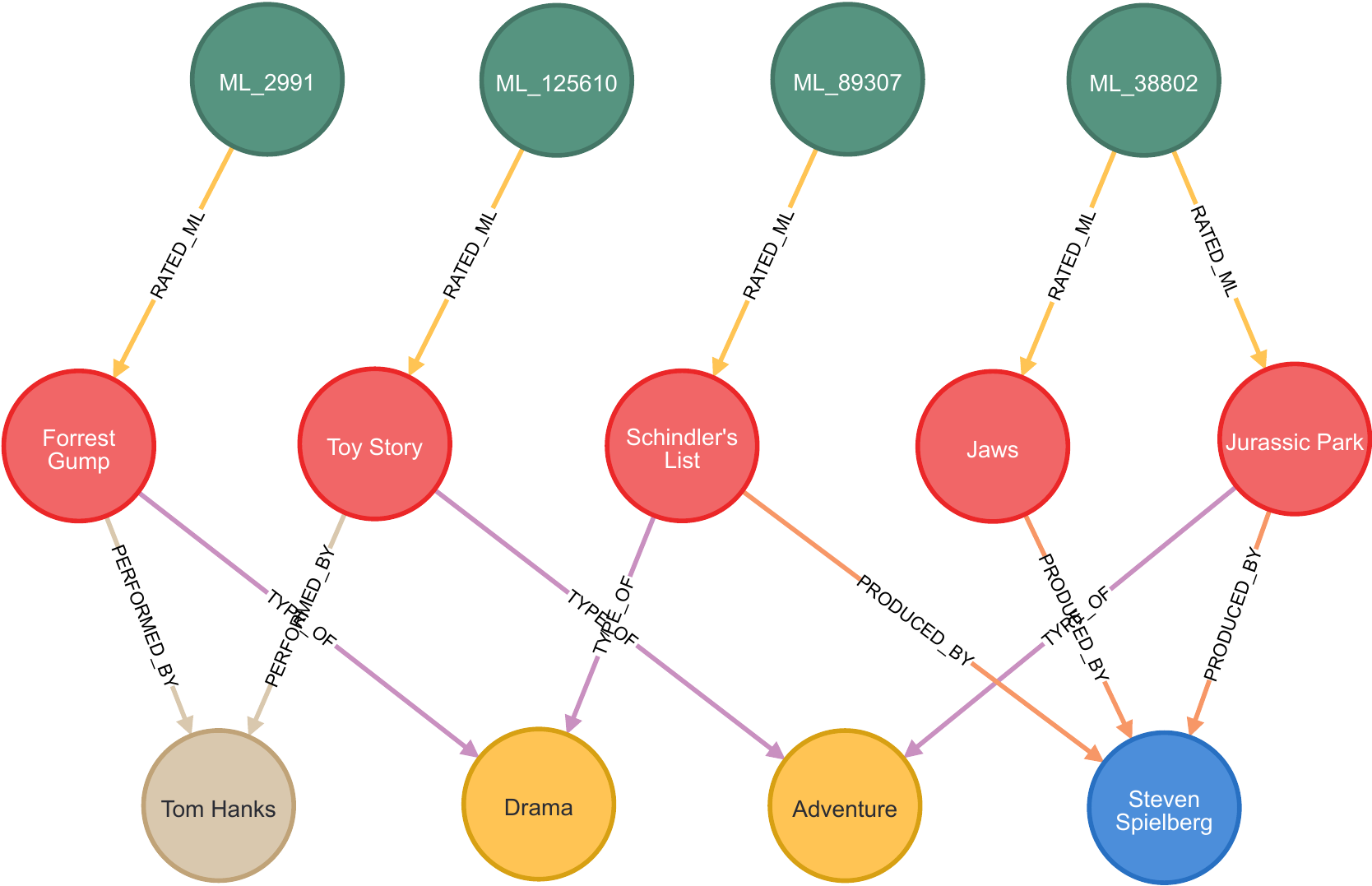}
  \caption{Integrating Yahoo!Movies into MovieLens on RecKG}
  \label{fig:integration_4.2}
\end{figure}

\subsection{Analysis of Integrated RecKG}
When integrating two different RecKG, the enhanced RecKG makes it easier to uncover concealed semantics that may be difficult to identify using the original KG alone. We will provide a detailed explanation of this in Figure \ref{fig:integration_4.2}.

The addition of new nodes broadens the semantic scope of the KG. As the performer attribute solely exists in the YM dataset, the performer nodes are added into the integrated RecKG, which is based on the ML dataset. Among these performer nodes, the `Tom Hanks' node is connected to the movie node `Forrest Gump' through the \textit{PERFORMED\_BY} relation. This connectivity implies that the user node `ML\_125610' from the ML dataset forms a three-hop neighbor relationship by following the path (\textit{ML\_125610 \(\xrightarrow{\text{RELATED\_ML}}\) Toy Story \(\xrightarrow{\text{PERFORMED\_BY}}\) Tom Hanks \(\xrightarrow{\text{PERFORMED}}\) Forrest Gump}), suggesting a potential preference for the movie node `Forrest Gump' by the user node `ML\_125610'. In practice, high-order connectivity is an important consideration in KG-based recommender systems, as it helps enhance the quality of recommendations \cite{kgat}.

Finally, it becomes possible to perceive preferences for specific side information of items. The inclusion of the producer node, originally from the YM dataset, into the ML dataset results in the integrated RecKG encompassing not only the performer node but also the producer node. This enables capturing latent preferences towards the movie node `Schindler's List' for the user node `ML\_38802'. This is attributed to the presence of two three-hop neighbor paths: (\textit{ML\_38802 \(\xrightarrow{\text{RELATED\_ML}}\) Jurassic Park \(\xrightarrow{\text{PRODUCED\_BY}}\) Steven Spielberg \(\xrightarrow{\text{PRODUCED}}\) Schindler's List}) and (\textit{ML\_38802 \(\xrightarrow{\text{RELATED\_ML}}\) Jaws \(\xrightarrow{\text{PRODUCED\_BY}}\) Steven Spielberg \(\xrightarrow{\text{PRODUCED}}\) Schindler's List}). These paths are established through the integration of the YM dataset into the ML dataset, where the YM dataset includes the director node 'Steven Spielberg,' which is not present in the ML dataset. Furthermore, it is worth noting that the movie node `Schindler's List' belongs to a completely different movie genre compared to `Jurassic Park' and `Jaws.' Since these movies have distinct genres, identifying the preference of the director node would have been challenging, as the ML dataset contains very little information and does not include the director node. In other words, due to the limited information available in the ML dataset, seizing a preference for the specific director node `Steven Spielberg' would have been challenging. As a result, it becomes possible to identify the preference of the user node `ML\_38802' for another movie directed by `Steven Spielberg', namely, `Schindler's List'. In summary, RecKG's expansion not only supports the revelation of item preferences but also preferences related to specific item information.

\subsection{Qualitative Evaluation}

\begin{table}
\caption{A qualitative evaluation between RecKG and other KGs for recommender systems.}
\resizebox{1.0\columnwidth}{!}{  
\setlength\tabcolsep{3.0pt}
\centering
\begin{tabular}{lcccc}
\toprule
\textbf{Characteristic} & \textbf{KGAT} \cite{kgat} & \textbf{KGCN} \cite{kgcn} & \textbf{MCCLK} \cite{mcclk} & \textbf{RecKG} \\
\midrule
\textbf{User Attribute}    &$\times$    &$\times$     &$\times$     &$\circ$\\
\textbf{Item Attribute}    &$\circ$     &$\circ$     &$\circ$     &$\circ$\\
\textbf{Interaction}   &$\triangle$     &$\triangle$     &$\triangle$     &$\circ$\\
\textbf{Consistency}   &$\times$     &$\times$     &$\times$     &$\circ$\\
\textbf{Interoperability}   &$\times$     &$\times$     &$\times$     &$\circ$\\
\bottomrule
\label{evaluation}
\end{tabular}
}
\begin{tablenotes}
\item[*] \fontsize{8}{9}\selectfont{* $\times$: Not considered, $\triangle$: Partially considered, $\circ$: Considered}
\end{tablenotes}
\end{table}

In this section, we highlight the interoperability of RecKG, demonstrating its ability to embrace various KG within the realm of recommender systems. To illustrate this, we compare RecKG with the KG formats employed in existing KG-based recommender system studies \cite{kgat, kgcn, mcclk}, as shown in Table \ref{evaluation}.

There are a total of 5 characteristics used for comparison: `User Attribute’, `Item Attribute’, `Interaction’, `Consistency’, and `Interoperability’.
The characteristics `User Attribute', `Item Attribute', and `Interaction' each indicate the presence or absence of user/item attributes and user-item interactions within the KG.
The `Consistency’ characteristic indicates whether or not the same naming is applied for the attributes according to their characteristics, irrespective of the dataset.
Finally, the `Interoperability' characteristic reflects the ability to represent and integrate most recommender system KG, ensuring both \textit{consistency} and \textit{diversity}.

Table \ref{evaluation} presents a comparative analysis of RecKG by examining the KG used in the KGAT \cite{kgat}, KGCN \cite{kgcn}, and MCCLK \cite{mcclk}.  
In these studies, `Item Attribute' is the only characteristic fully addressed, whereas the `User Attribute' is not considered.
This is mainly because each study focuses solely on KG derived from item attributes and user-item interactions.
Regarding `Interaction,' various types of user-item interactions, including purchase, add-to-cart, and add-to-favorites, are all treated uniformly. 
Although most studies treat the multi-typed user-item interactions as the same, each interaction type carries a distinct meaning \cite{user_behavior, xia2020multiplex}. 
Therefore, taking these diverse interactions into account can improve the performance of KG-based recommender systems.
RecKG, in contrast, addresses the aspect of \textit{diversity} by containing item attributes, user attributes, and various types of user-item interactions. 
Additionally, RecKG facilitates consistent naming across different datasets, depending on the type of the attribute, due to its focus on the `Consistency'. 
Consequently, RecKG is the only KG that incorporates the `Interoperability' characteristic, ensuring both \textit{diversity} and \textit{consistency}. Through a qualitative comparison with other studies, it can be assessed that RecKG covers most KG in the recommender system field and takes interoperability into account.

\section{Conclusion}


In this study, we introduce RecKG, a standardized Knowledge Graph (KG) for achieving interoperability in recommender systems, addressing the challenge of diverse representations. RecKG focuses on \textit{consistency} and \textit{diversity}, employing a comprehensive naming scheme for user and item attributes and minimizing missing attributes. Demonstrated through case studies and a graph database application, RecKG proves effective.

Future plans involve researching KG integration across domains using RecKG and experimenting with KG-based recommender systems within the integrated KG. We anticipate RecKG to standardize KGs, offering a comprehensive overview for future KG-based recommender system research.

\section*{Acknowledgement}
This work was partly supported the National Research Foundation of Korea (NRF) grant funded by the Korea government (MSIT) (NRF-2022R1C1C1012408) and Institute of Information \& communications Technology Planning \& Evaluation (IITP) grants funded by the Korea government (MSIT) (No.RS-2022-00155915, Artificial Intelligence Convergence Innovation Human Resources Development (Inha University) and No.2022-0-00448, Deep Total Recall: Continual Learning for Human-Like Recall of Artificial Neural Networks).

\balance
\bibliographystyle{ACM-Reference-Format}
\bibliography{RecKG_main} 

\end{document}